\documentclass[amsmath,prb,twocolumn]{revtex4}
\usepackage{graphicx}

\begin{document}

\title{Free Energy Landscape of a Protein-Like Chain in a Fluid with Discontinuous Potentials}

\author{Hanif Bayat Movahed$^1$}\email{hbayat@chem.utoronto.ca}
\author{Ramses van Zon$^{1,2}$}
\author{Jeremy Schofield$^1$}

\affiliation{%
  $^1$Chemical Physics Theory Group, Department of Chemistry,
  University of Toronto, 80 St.\ George Street, Toronto, Ontario M5S
  3H6, Canada\\ 
  $^2$SciNet High Performance Computing Consortium, University of
  Toronto, 256 McCaul Street, Toronto, Ontario M5T 1W5, Canada 
}

\date{December 15, 2011}

\begin{abstract}
  The free energy landscape of a protein-like chain in a fluid was
  studied by combining discontinuous molecular dynamics and parallel
  tempering. The model protein is a repeating sequence of four
  different beads, with interactions mimicking those in real proteins.
  Neighbor distances and angles are restricted to physical ranges and
  one out of the four kinds of beads can form hydrogen bonds with each
  other, except if they are too close in the chain.  In contrast to
  earlier studies of this model, an explicit square-well solvent is
  included.  Beads that can form intra-chain hydrogen bonds, can also
  form (weaker) hydrogen bonds with solvent molecules, while other
  beads are insoluble.  By categorizing the protein configurations
  according to their intra-chain bonds, one can distinguish unfolded,
  helical, and collapsed helical structures.  Simulations for chains
  of 15, 20 and 25 beads show that at low temperatures, the most
  likely structures are helical or collapsed helical, despite the low
  entropy of these structures.  The temperature at which helical
  structures become dominant is higher than in the absence of a
  solvent.  The cooperative effect of the solvent is attributed to the
  presence of hydrophobic beads.  A phase transition of the solvent
  prevented the simulations of the 20-bead and 25-bead chains of
  reaching low enough temperatures to confirm whether the free energy
  landscape is funnel-shaped, although the results do not contradict
  that possibility.
\end{abstract}

\maketitle

\section{Introduction}
\label{sec:Intro}

Proteins have a natural tendency to find a unique folded
structure. Understanding how proteins fold, has been a very
challenging question in physics, chemistry and
biology\cite{shakh:2,Science:63}, which remains largely an open
problem.

One of the contentious issues is why folding occurs so fast. According
to Levinthal, the time required to explore all conformations of an
average protein is too long to find the global free energy minimum on
realistic time scales.\cite{Levinthal:53,Wolynes:45,Goldstein:55} This
seems to clash with the idea that the native configuration of a
protein is associated with the global minimum its Gibbs free
energy.\cite{Wenzel:10, Anfinsen:12} To resolve these two apparently
conflicting views, it is useful to think of folding as occurring on a
\emph{free energy landscape}. The free energy landscape is the form of
the free energy as a function of protein
conformation.\cite{Wolynes:45} The paradox would be resolved if the
landscape has a \emph{funnel} shape. The system could then slide down
the funnel towards the native state through configurations of
increasingly lower free energy --- much faster than finding the native
state through a random walk --- but without needing a single, specific
kinetic pathway.

There is, furthermore, no consensus as how significant the role of the
solvent is for protein folding, or what interaction or set of
interactions play the main role.\cite{Woolfson:32} Some researchers
proposed that folding is a balance between entropy versus
enthalpy-dominated hydration.\cite{Athawale:13,Garde:14} On the other
hand, there have been experiments that have shown that a protein can
fold into its native configuration with apparently negligible solvent
ordering effects.\cite{Woolfson:32,Soundararajan:73} In many of the
theoretical and experimental studies of protein folding, the proteins
have been in the absence of a fluid.\cite{Rhee:33} However in nature,
and therefore in many experimental studies, folding happens in the
presence of a fluid environment.

A simple model of a protein without a solvent was studied in
Ref.~\onlinecite{hanif:34}; we will refer to this work as \textbf{I}
below. The simple models in \textbf{I} were used to capture the basic
behavior of proteins in a reasonable computational time.  This was
accomplished by using discontinuous potentials for the attractive and
repulsive interaction, as step and shoulder potentials
respectively. In addition, here, the protein-like chain is surrounded
by an explicit solvent environment. As in \textbf{I}, the
investigation of the free energy landscape was done using a Hybrid
Monte Carlo (HMC) method, where HMC is implemented as a combination of
the Monte Carlo and the Discontinuous Molecular Dynamics (DMD)
method. The Parallel Tempering (PT) method \cite{Swendsen:6, Greyer:7,
  whittington:18} is used for the Monte Carlo part to avoid getting
trapped in local free energy minima and to increase the speed of phase
space exploration.\cite{Earl:8} The PT method allows to generate
configurations according to the canonical ensemble.

Our earlier study resulted in a free energy landscape for the
protein-like chain in the absence of any fluid. Using a family of
simple protein models consisting of a periodic sequence of four
different kinds of bead, these protein-like chains exhibited a
secondary alpha helix structure in their folded states, and allowed a
natural definition of a configuration by considering which beads are
bonded.  Relative configurational free energies at different
temperatures were determined from relative populations at those
temperatures.

In the absence of any fluid it could be demonstrated that the energy
landscape is rugged at different temperatures and at low enough
temperature has a very deep funnel shaped valley especially for the
smaller chains. In other words, one structure has a much lower free
energy than other structures, so that its probability becomes very
high. At the same time, neighboring structures should still have a low
free energy, so that folding could be driven through configurations of
progressively lower free energy. The much lower free energy of the
global minimum was confirmed by the fact that, for chains smaller than
30 beads, the probability of observing the most common configuration
was almost one at low enough temperatures.

In this paper we investigate the energy landscape of the protein-like
chain in the presence of a fluid, to be compared with the results of
\textbf{I}. For this purpose, solvent particles are added to the
system with the protein-like chain. As was the case for the
intra-chain interactions, the interaction between the solvent
particles as well as the solvent particles with the chain beads are
defined by discontinuous potentials.

The paper is structured as follows.  In
Sec.~\ref{sec:Protein-likeChain}, the models for the protein-like
chain and the solvent and their parameters are described.  In
Sec.~\ref{sec:Simulation-Techniques}, the simulation techniques are
introduced.  In Sec.~\ref{sec:Results}, the efficiency of the
simulations and the complications due to a solvent phase transition
are discussed, and the results for the most populated structures of
the systems containing 15, 20 and 25 beads chain are presented.
Finally, in Sec.~\ref{sec:concl}, the conclusions will be given.

\section{The system}

\subsection{The protein-like chain model}
\label{sec:Protein-likeChain}

The system consists of two parts: the protein and the solvent. The
protein part of the model is the same as model B from \textbf{I}. It
is a \emph{beads on a string} model in which each bead represents one
amino acid or residue. The chain consists of a repeated sequence of
four different kinds of beads. The interactions between these beads
are designed to mimic the secondary structure of an alpha helix.
While in the previous work, chain lengths $\ell$ varied from 15 to 35,
in this work we consider the cases $\ell=15,20$ and 25, because the
presence of the solvent reduces the chain length that can be studied
without becoming computationally prohibitive, and because, for these
values of $\ell$, the free energy landscape of the solvent-less model
was found to have a funnel shape.\footnote{Other studies indicate that
short chains containing 6, 8 or 12 monomers are too short to capture
compact states, while somewhat longer chains with 25 monomers can
capture folded helical states.\cite{Athawale:13}} To make contact
with real proteins, and because there are too many parameters to form
unique reduced units, physical units are used in the definition of the
model, although these should not be taken too literally: we only aim
to set these to the right order of magnitude to mimic real
proteins. In particular, lengths will be expressed in \AA ngstr\"oms,
energies in kJ/mol and masses in atomic mass units. In the model of
the protein-like chain, the mass of each residue is $m_p=$120 amu, or
$2\cdot10^{-25}$ kg and five kinds of interaction are defined. In the
first two kinds of interactions, distances between nearest and
next-nearest neighbor beads are confined to a specific ranges by an
infinite square-well potential similar to Bellemans' bonds
model.\cite{Bellemans:26} (a)~In the first kind of interaction, which
mimics covalent bonds for nearest neighbors, the distances are
restricted to lie between 3.84~\AA\ to 4.48~\AA. (b)~The second kind
of interaction is a next-nearest neighbors infinite square-well
potential with a range from 5.44~\AA\ to 6.40~\AA\ to represent the
angle vibration between 75$^\circ$ and 112$^\circ$.  (c) The third
kind of interaction are hydrogen bonds within the chain, which are
modeled by an attractive square-well potential with a range from
4.64~\AA\ to 5.76~\AA\ and a depth of $\epsilon= 20$ kJ/mol. These
only act between beads $i$ and $i+4n$, where $i=4k+2$ (k is an integer
number) and $n$ cannot be 2 or 3.  (d)~A repulsion in the form of a
shoulder potential acts between beads $1+4k$ and $4k'$, where $k$ and
$k'$ are integers and $k \neq k'$. The range of the shoulder is from
4.64~\AA\ to 7.36~\AA , while the height is $0.9
\epsilon$. (e)~Finally, all other bead pairs for which neither a
covalent, hydrogen bonds or repulsive interactions are defined, feel a
hard sphere interaction with a diameter of 4.6~\AA.

\subsection{The solvent model}

The solvent consists of $N$ molecules in a fixed volume~$V$ which
interact via a square-well potential. The square-well fluid has been
studied extensively.\cite{ErringtonRf3:35, OreaRf4:36, Schollref5:37,
  Escamillaref6:38, Orkoulas:40, Elliott:41, Rio:42, Vega:43, Lang:44}
To be able to compare to the previous studies, a popular set of
parameters has been used where $\sigma$ and $\sigma'$, representing
the inner and outer points of discontinuity of the potential well,
satisfy $\frac{\sigma'}{\sigma}=1.5$. In particular, $\sigma$ and
$\sigma'$ are chosen to be 4.16~\AA\ and 6.24~\AA, respectively, and
the potential depth for the square-well interaction between the fluid
particles, $\epsilon_l$, is defined as $\frac{0.35}{1.5}\epsilon
\simeq 0.23 \epsilon$, or 4.7 kJ/mol.

The mass of each fluid particle is chosen as $m_l=0.15m_p$, where
$m_l$ and $m_p$ are the masses of fluid particles and chain beads
respectively. This choice makes the fluid particles much lighter than
the chain beads. In physical units, the solvent particle mass is very
close to that of a water molecule, i.e., 18 amu.\footnote{The masses
of the constituents do not influence the thermodynamic properties
such a free energies, but solvent molecules that are more massive
would influence the sampling efficiency of the simulations.}

The solvent and the chain interact as follows. The solvent particles
can make hydrogen bonds with the chain beads $i=4k+2$, where $k$ is a
positive integer number, with a potential depth of $\epsilon_l$. The
interaction range is the same as the hydrogen bonds between the chain
beads (i.e., $\sigma_1$ and $\sigma_2$ are 4.64~\AA\ and 5.76~\AA,
respectively). Hence, the same beads that are involved in making bonds
inside the protein-like chain are involved in making hydrogen bonds
with the solvent particles. Other chain beads have a hard sphere
repulsive interaction with the solvent particles. The hard sphere
interaction range is set to a relatively large value of 6.4~\AA (1.54
$\sigma$) to mimic the hydrophobicity of amino acids.

The simulation occurs in a cubic box of size $L\times L\times L$ that
contains $N$ solvent particles and one protein-like chain. To minimize
finite-size effects, periodic boundary conditions are used. To avoid
boundary artifacts, $L$ should be chosen large enough to allow the
protein-like chain being stretched without the last two end beads of
the chain affecting each other (either directly or though solvent
mediated interactions). The maximum observed value for the end-to-end
vector in the previous study in \textbf{I} is considered the worst
case scenario, and the value for $L$ was chosen to be comfortably
larger.  Because of the next-nearest neighbor distance restriction,
the maximum end-to-end distance can be determined analytically from
the model's definition. The values used for $L$ are roughly
10~\AA\ larger than the theoretical maximum end-to-end distance, which
is itself substantially larger than the observed worst-case end-to-end
distance in the absence of a fluid. For example, for the 25-bead chain
the maximum observed value for the end-to-end vector is 64~\AA, while
theoretical calculation shows a maximum possible value of
76.8~\AA. The value used for $L$ is 88.0~\AA\ (21.15 $\sigma$), which
is 24~\AA\ larger than the maximum observed value in the simulation
runs and 11.2~\AA\ larger than the theoretical maximum
value. Following a similar reasoning, for the $\ell=$15 and 20-bead
chains, the values of $L$ are set to 54.4~\AA\ (13.08 $\sigma$) and
72.0~\AA\ (17.31 $\sigma$), respectively.

With the total volume of the simulation box determined, one sets the
number of particles $N$ such that the solvent has the required density
$\rho^*=\rho\sigma^3 $, where $\rho=\frac{N}{V_l}$, and $V_l$, the
effective free volume that fluid particles can occupy, equals
$L^3-V_{excl}$, where $V_{excl}$ is the approximate excluded volume of
the chain. To calculate the approximate excluded volume of the chain,
it is assumed that the protein-like chain lies completely straight and
the distance between two neighboring beads is 4.16~\AA, which is the
mid point of vibrating distance of protein-like beads.  Then the
volume of the cylinder around this chain, in which no other bead can
exist, is considered as the excluded volume. The reduced density
$\rho^*$ was chosen to be 0.5 and consequently, $N$ are 1066, 2522 and
4644 for the $\ell=$15, 20 and 25 bead chains, respectively.

It will be convenient to introduce a dimensionless temperature
scale. The reduced temperature is defined as $T^*=k_b T / \epsilon$,
however another reduced temperature, $T^*_l$, is defined using the
potential depth of the fluid particles square-well interactions to
make the comparison easier with earlier studies of the phase diagram
of this type of fluid. Hence, $ T^*_l=k_b T / \epsilon_l$, where
$T^*_l=(\epsilon / \epsilon_l)T^*=(1.5 / 0.35)T^* \simeq
4.29T^*$. $\beta^*$ and $\beta^*_l$ are defined as the inverse
functions of $T^*$ and $T^*_l$ respectively. Note that $T^*=1.0$
corresponds to 2400K, while $T_l^*=1.0$ corresponds to 560K and $T_l^*
\simeq 0.5$ is roughly room temperature.

\subsection{Definition of configurations}
\label{sec:DefinitionConfigurations}

To determine the free energies of different configurations, one first
has to decide how configuration are defined. Here only intra-chain
interactions are counted to identify a configuration. Since there are
additional interactions (solvent-chain and solvent-solvent), a
configuration does not have a unique energy within this model, in
contrast to the case in \textbf{I}, which had no explicit solvent, and
where, as a result, the energy of a configuration was constant. As in
\textbf{I}, a configuration is represented by a string of alphabetical
pairs. For example, BF represent a configuration with one bond between
beads 2 and~6, and BF FJ JN represents a configuration with three
bonds, between beads 2 and 6, 6 and 10, and 10 and 14.

By identifying configurations regardless of the solvent particles, the
free energies that will be found are averaged over those degrees of
freedom, in line with the ideas of Refs.~\onlinecite{Wolynes:45},
\onlinecite{Wolynes:22}, and \onlinecite{Young:46}. The free energy
values are further coarse-grained in the sense that they are not a
function of all the positions of the atoms in the chain, but they are
a function of the absence or presence of bonds.

Note that instead of the free energy $F_c$ we will often report the
population, or frequency of observing, of configurations $c$, which
will be denoted by $f_{obs,c}$. These two quantities are directly
related by
\[
   \frac{f_{obs,a}}{f_{obs,b}} = e^{-\beta[F_a-F_b]}. 
\]
where $a$ and $b$ are two configurations. In other words,
$F_c=const-k_BT\ln f_{obs,c}$. Thus, low populations corresponds to
high configurational free energy and populations near 100\% correspond
to the highest possible configurational free energy.

\section{Simulation Techniques}
\label{sec:Simulation-Techniques}

The simulation uses a combination of DMD and PT, in which the simulated system consists of a number of replicated protein-like chains inside a solvent.\cite{hanif:34, Rappa:4, Hernandez:9} All replicas evolve for a fixed amount of time using DMD, after which some of the replicas exchange their temperatures. The velocities of the solvent and the bead particles of all the replicas are drawn from the Maxwell-Boltzmann distribution both initially and at the end of any replica exchange event. Since the velocities of all replicas are being updated periodically using the Maxwell-Boltzmann distribution and the DMD dynamics is reversible and preserves phase space volume, all necessary conditions for generating a state with canonical distribution are satisfied.\cite{Duane:30}

Because the number of solvent particles required to avoid boundary
effects scales with the third power of number of beads in the chain,
as this number increases, exploring the energy landscape becomes more
challenging, involving thousands of particles. To address this
computationally demanding issue, we developed a parallel program using
the Message Passing Interface technique.\cite{MPI:47} In the parallel
program, each replica runs on one processor. Communication only occurs
at the replica exchange event, at which point the energy values of the
replicas are sent to the master processor, which determines whether a
temperature exchange should take places, and then sends each replica
its updated temperature (which can be the same as its earlier
temperature). The different processes for each replica then
independently draw the velocities from this new (or old) temperature,
and from there, start another DMD run. The process of drawing
velocities, DMD dynamics, and PT exchange moves is repeated until
enough independent statistics on the population of different
configurations is gathered.

\section{Results}
\label{sec:Results}

\begin{figure}[t]
  \centering
  \includegraphics[angle=-90,width=\columnwidth]{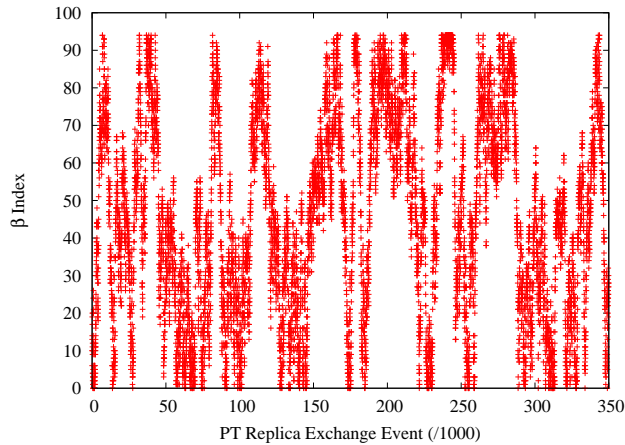}
  \caption{Proper temperature dynamics for one of 95 replicas for $\ell=15$.}
  \label{fig:Feb25-s15-350-R10}
\end{figure}

\subsection{Parallel tempering efficiency}
\label{sec:healthyPT}

In the current context, we call the simulation efficient if it
generates many independent configurations in an unbiased way in a
given simulated time period. For instance, since the PT simulations
can be seen as replicas moving from temperature to temperature while
they change their configurations, if a certain replica gets stuck in a
certain range of temperatures, the sampling would be biased. To obtain
good sampling, one should tune the number of used replicas, the
temperature difference between successive temperatures $\Delta \beta$,
and the duration of the simulated time period between consecutive
replica exchange events, the so-called \emph{PT update period}. These
parameters have a strong effect on the efficiency of dynamics. Since
decreasing the PT update period may cause the replicas to explore a
smaller part of the configurational space, there is an optimum value
for this parameter for a fixed computational cost, which has to be
found by trial and error. A key concept to assess the efficiency of a
PT simulation is a PT cycle, which is the simulated time for the
replica to travel between the minimum and maximum temperatures and
back.\cite{Sheldon:84} For efficient sampling, several cycles should
be observed in one run. The value for the PT update period which leads
to the largest number of PT cycles is different for various lengths,
$\ell$, of the chain under consideration. However, the number of
interaction events during each PT update period for 15, 20 and 25
beads chains are similar. This provides a good initial estimate for
the optimum value of the PT update period of the larger systems based
on the results of smaller systems, and facilitates the trial and error
process.

In principle, increasing the number of replicas would make it possible
to study any range of temperatures. However, it was found that when
the PT system contains a large number of replicas, some of the
replicas may not move very well among the full range of temperatures
during one PT cycle. This can lead to a prohibitively inefficient PT
dynamics. In addition, it was found that the presence of a phase
transition in the solvent model reduces the range of temperatures that
can be studied (see Sec.~\ref{phaseofthesolvent}).

\begin{figure}[t]
  \centering
  \includegraphics[angle=-90,width=\columnwidth]{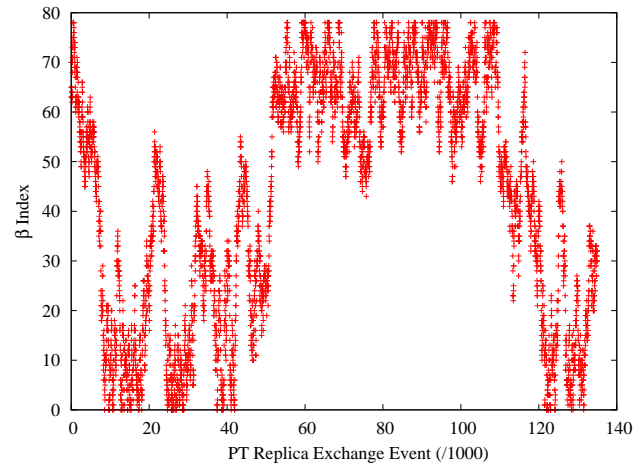}
  \caption{The effect of phase transition, which happens around
    $\beta$ index 50, on the PT dynamics for one of 79 replicas for
    $\ell=15$. }
  \label{fig:s15-79-135-R70}
\end{figure}

As an example, Fig.~\ref{fig:Feb25-s15-350-R10} shows proper dynamics
for the 15-bead chain in which the temperature range
$T_l^*=[0.76,2.5]$ is investigated by 95 replicas. For this case the
inverse temperature difference $\Delta \beta_l^*$ for the 10 replicas
with the highest temperatures is 0.012 and in the next 60 replicas the
$\Delta \beta_l^*$ decreases linearly to 0.008 and then it remains
constant. The PT update period for this case is 0.8 ps. The motivation
for using a varying $\Delta \beta$ can be found in~\textbf{I}. Plots
like the one in Fig.~\ref{fig:Feb25-s15-350-R10} are a helpful tool in
checking for poor sampling. The example in
Fig.~\ref{fig:s15-79-135-R70} shows what such a plot looks like for a
poorly behaving PT simulation in which the temperature range
$T_l^*=[0.82,2.5]$ is investigated using 79 replicas. For this case,
the PT update period is 2~ps, which is 2.5 times larger than the
previous case in Fig.~\ref{fig:Feb25-s15-350-R10}. $\Delta\beta_l^*$
for the highest 30 temperatures is 0.012, and then for the next 40
temperatures the $\Delta\beta_l^*$ decreases linearly to $0.008$ and
then $\Delta\beta$ remains constant for the rest of temperatures.

\begin{figure*}
  \centering
  (a)\hspace{0.5\textwidth}(b)\\
  \includegraphics[angle=-90,width=0.49\textwidth]{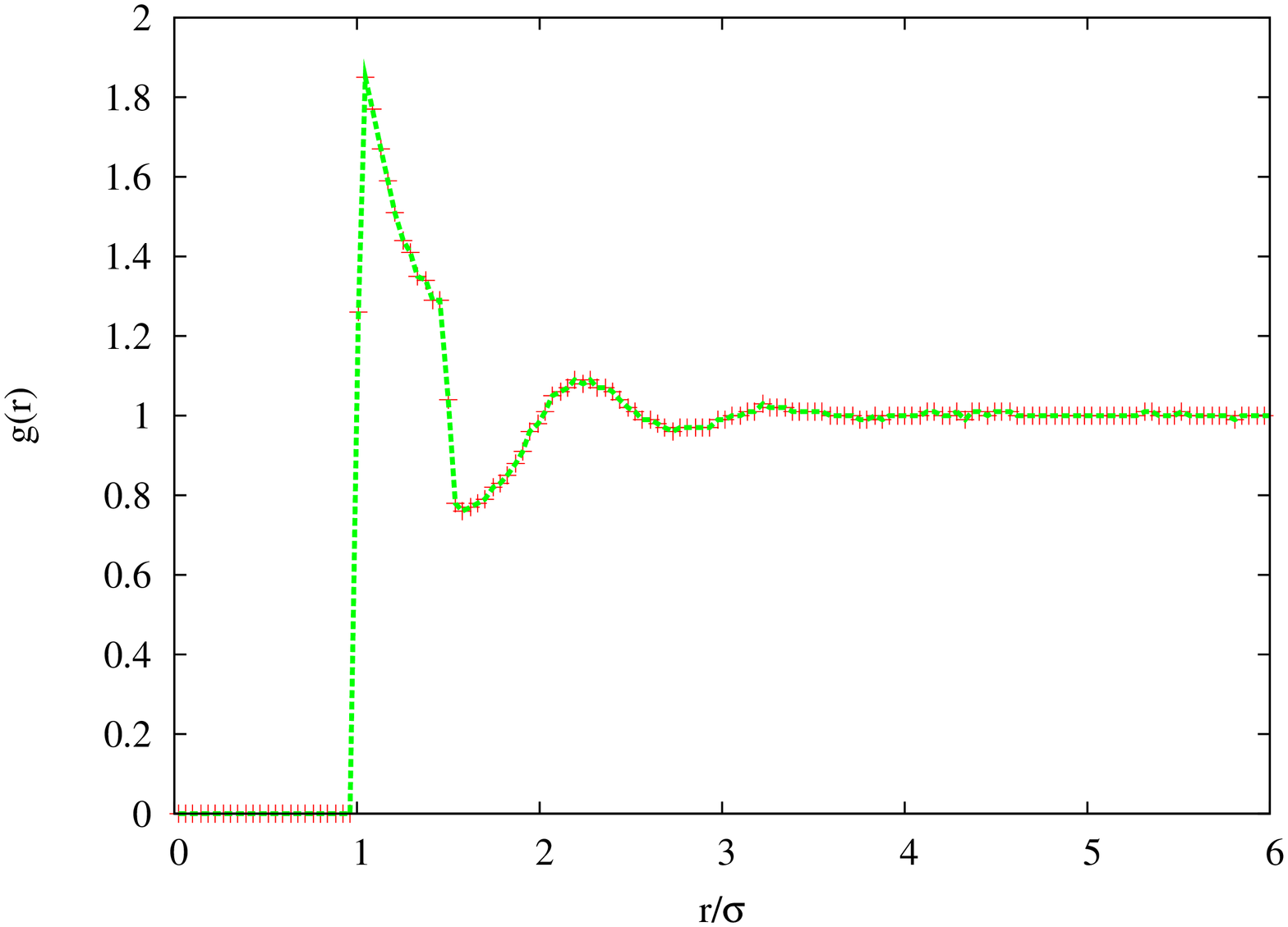}
  \includegraphics[angle=-90,width=0.49\textwidth]{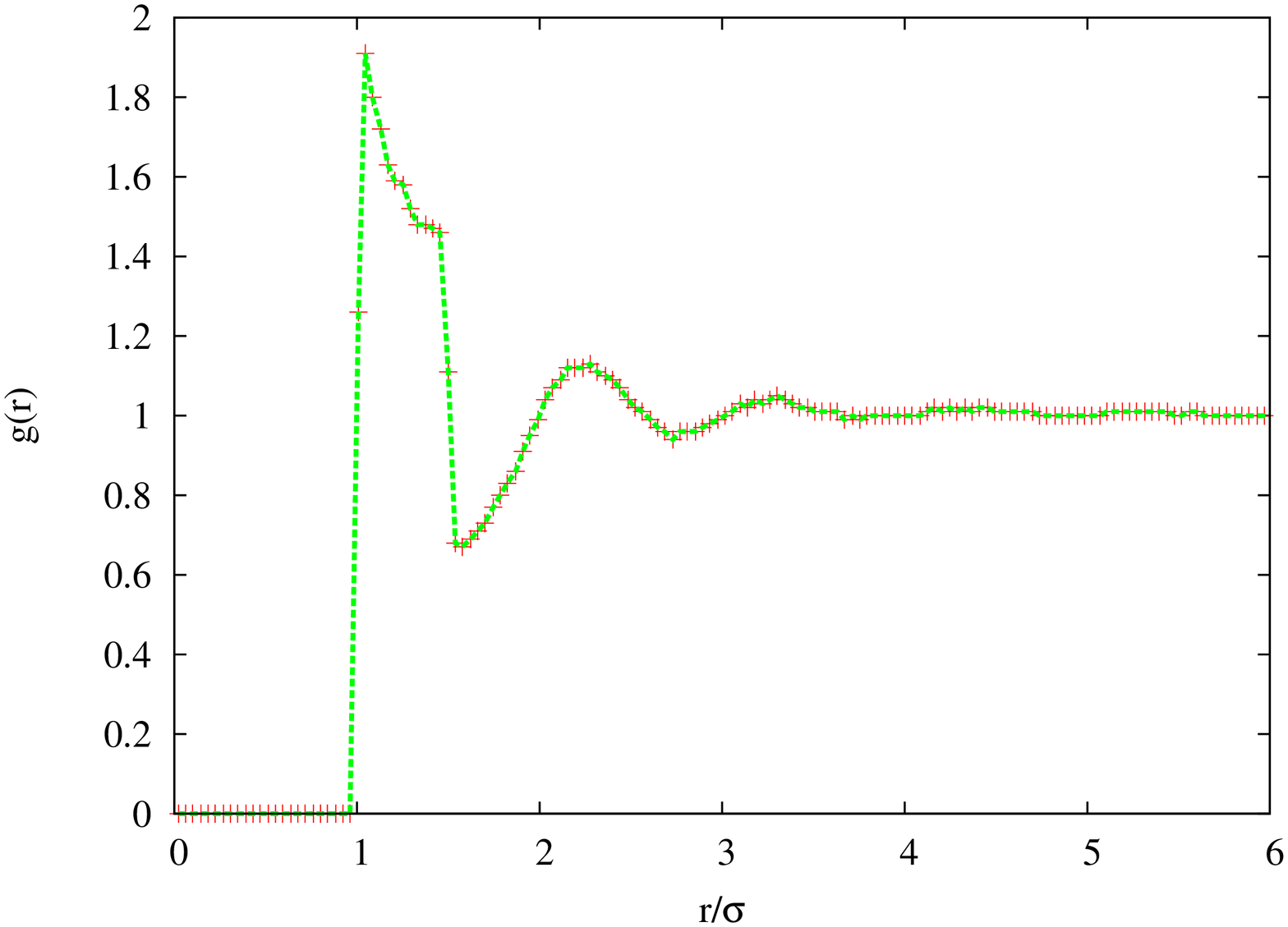}\\\ \\
  (c)\hspace{0.5\textwidth}(d)\\
  \includegraphics[angle=-90,width=0.49\textwidth]{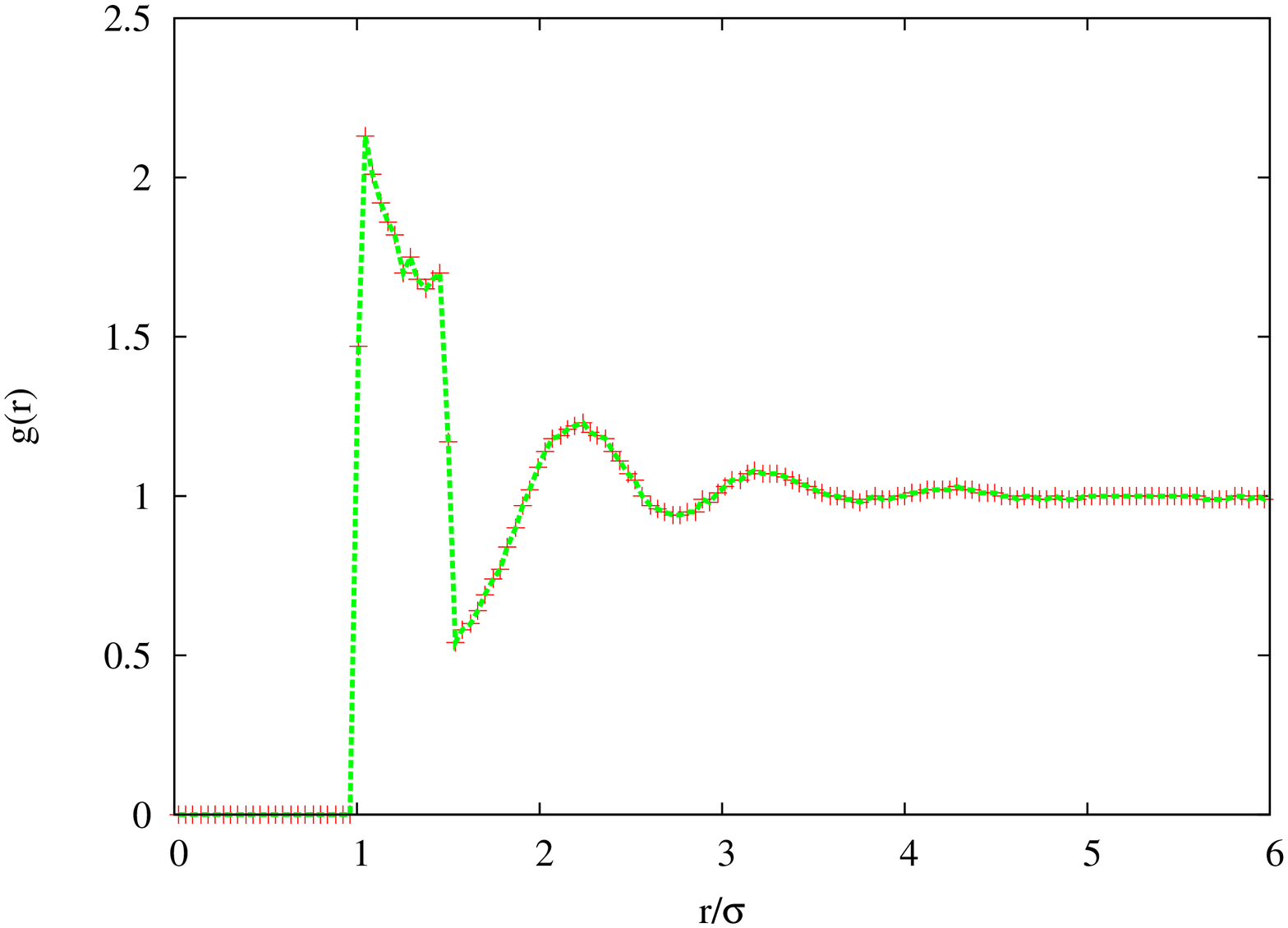}
  \includegraphics[angle=-90,width=0.49\textwidth]{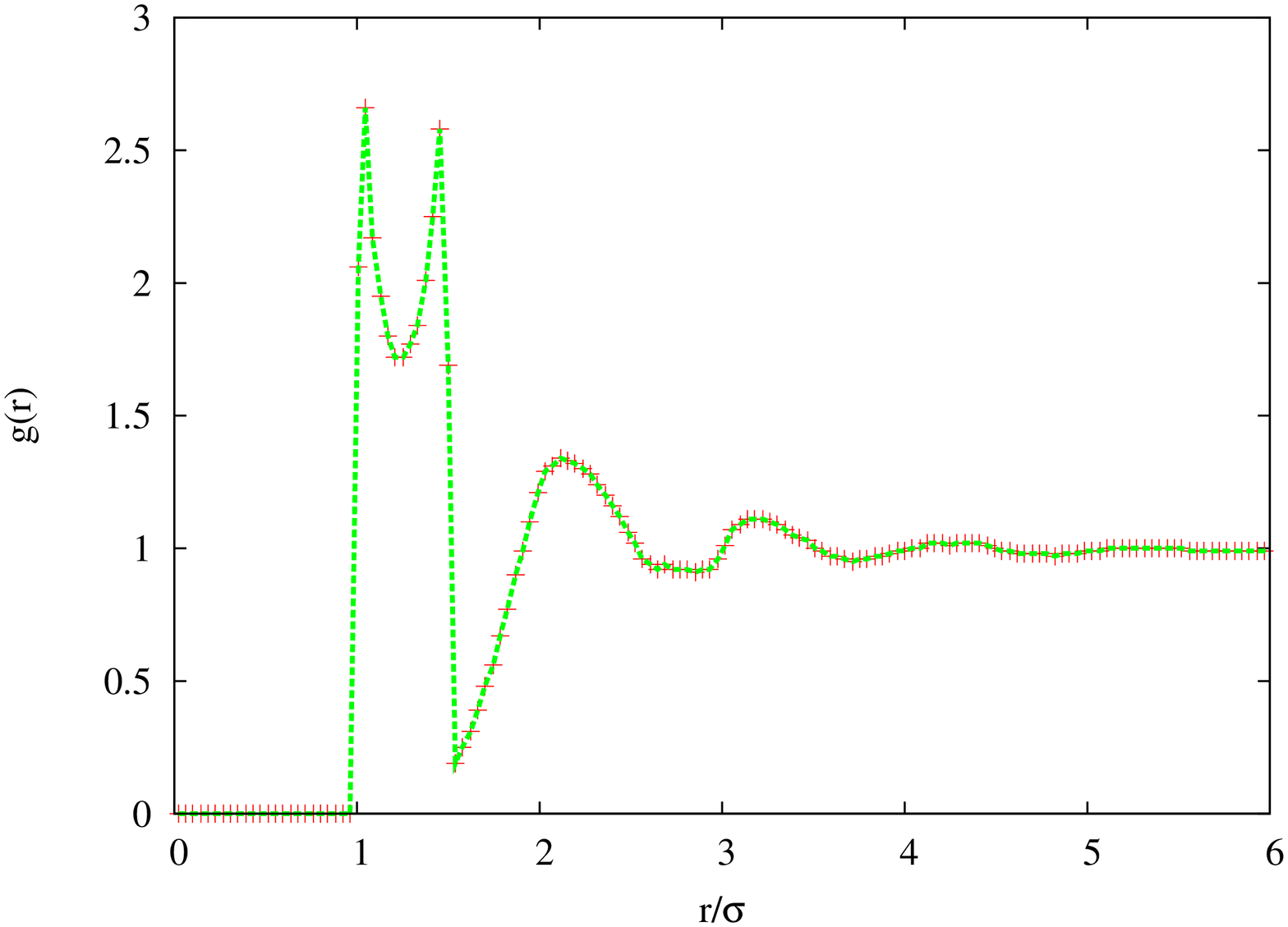}
  \caption{Radial distribution function for the square-well fluid at
    different temperatures $T_l^*$: (a) $2.0$, (b) $1.25$, (c) $0.83$,
    and (d) $0.31$. Lines are drawn to guide the eye.}
  \label{fig:allRDF}
\end{figure*}

\subsection{Phase of the solvent}
\label{phaseofthesolvent}

Figure~\ref{fig:s15-79-135-R70} shows an apparent barrier in the PT
dynamics at a specific temperature. In other words, replicas have a
strong tendency to stay above, or below, that specific temperature,
and rarely cross it. While choosing proper parameters that optimize
the efficiency of the PT method can improve sampling in the vicinity
the temperature barrier, the efficiency of the sampling drops
significantly for larger systems (i.e. for $\ell=20$ and
$\ell=25$). It turns out that the apparent barrier is related to the
phase of the solvent.

The highest temperature in the simulations was $T_l^*=2.5$ for all
chain lengths, while the lowest values for $T_l^*$ were 0.76, 1.05 and
1.22 for the 15-bead, 20-bead and 25-bead systems, respectively. The
square-well model for the solvent has been studied
extensively\cite{ErringtonRf3:35, OreaRf4:36, Schollref5:37,
  Escamillaref6:38, Orkoulas:40, Elliott:41, Rio:42} and for the model
used here with $\rho^* = 0.5$ and $\lambda=\sigma'/\sigma=1.5$, the
critical reduced temperature, $T_c^*$, for the solvent is predicted to
be 1.2172~\cite{ErringtonRf3:35}, 1.210 (in Ornstein-Zernike
approximation)~\cite{Schollref5:37}, 1.3603 (using an analytical
equation of state based on a perturbation
theory)~\cite{Schollref5:37}, 1.226~\cite{Escamillaref6:38},
1.2180~\cite{Orkoulas:40}, 1.27~\cite{Elliott:41} and
1.218~\cite{Rio:42}. Most of the previous
studies~\cite{ErringtonRf3:35, OreaRf4:36, Schollref5:37, Vega:43},
predict a vapor-liquid coexistence line to be crossed somewhere
between $T_l^*=1.0$ and $T_l^*=1.2$ for $\rho^*=0.5$ and
$\lambda=1.5$. The simulated systems are far from a real thermodynamic
system so one expect difficulties in observing this
transition. Furthermore, finite-size effects may shift the apparent
critical temperature.

As a first check to confirm the fluid-like character of the solvent
model, the radial distribution function (RDF) of the solvent was
studied for four different temperatures. These are plotted in
Fig.~\ref{fig:allRDF}. Due to the two discontinuities in the solvent
interaction potential at $\sigma$ and $\sigma'$, respectively, the
radial distribution function is relatively high between these two
points. For $T_l^*=2.0$, the RDF graph~\ref{fig:allRDF}(a) is very
similar to what was found for this model in the earlier studies (3rd
graph in Fig. 2 in Ref.~\onlinecite{Escamillaref6:38}). At this
temperature, fluid-like long range correlation can be seen
already. The RDF for $T_l^*=1.25$, Fig.~\ref{fig:allRDF}(b), and that
for $T_l^*=0.83$, Fig.~\ref{fig:allRDF}(c), look like those of a
typical fluid with more pronounced peaks than the high temperature
RDF. At relatively low temperatures of $T^*=0.31$, as in
Fig.~\ref{fig:allRDF}(d), the onset of short range structural peaks
may be showing itself in the first two peaks, while other peaks show a
fluid-like behavior, but there is no clear sign of a phase transition.

\begin{figure}[t]
  \centering
  \includegraphics[angle=-90, width=\columnwidth]{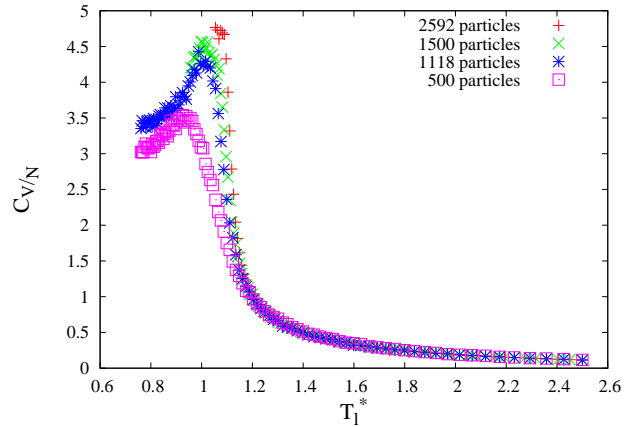}
  \caption{Heat capacity of the square-well solvent per particle
    vs. the liquid reduced temperature for $N=$ 500, 1118, 1500 and
    2592.}
  \label{fig:HCaverage}
\end{figure}

\begin{figure}[t]
  \centering
  \includegraphics[angle=-90,width=\columnwidth]{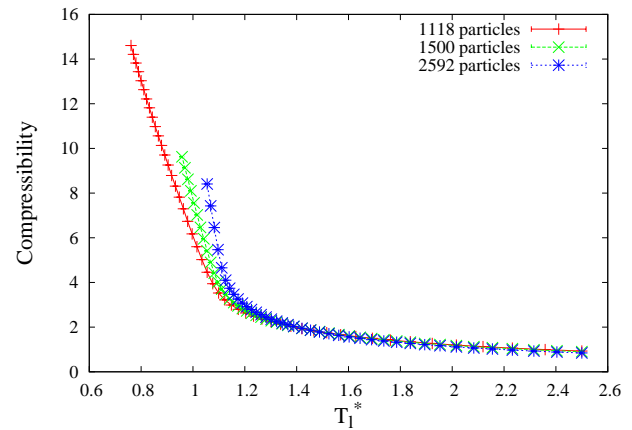}
  \caption{Compressibility of the square-well solvent vs. the liquid
    reduced temperature for $N=$ 1118, 1500 and 2592.}
  \label{fig:densities}
\end{figure}

RDFs are, however, not a very good indicator of a phase
transition. Better indicators are the heat capacity $C_v$ and the
compressibility $\kappa$, which are second derivatives of the free
energy. $C_v$ can be measured from the fluctuations in energy, while
$\kappa$ can be estimated from fluctuations in local density. For the
latter, the system is divided into several boxes and the densities in
each box and the standard deviation of the local density are
calculated. Numerical estimates for the head capacity and
compressibility in the canonical ensemble are plotted in
Figs.~\ref{fig:HCaverage} and \ref{fig:densities}. The range of
studied temperatures was clearly sufficient to observe the effects of
a phase transition for smaller systems. This phase transition occurs
at a temperature that is very close to the temperatures at which other
studies predict the liquid-vapor coexistence line for this density.

Fig.~\ref{fig:HCaverage} shows that the average heat capacity per
solvent particle increases with increasing system size at the phase
transition point. This suggests that for infinitely large systems, the
heat capacity might diverge at the phase transition. To better
understand the phase transition and its order, a further study would
be required which lies outside the scope of this paper.

In Fig.~\ref{fig:densities}, the variation of the compressibility
vs. temperature shows similar behavior to that observed for the heat
capacity. By increasing the system size, the compressibility seems to
diverge to infinity around the same point where the heat capacity
diverges. This confirms that there is a phase transition at this
point.

While these results are for a pure solvent system, our studies
revealed that there is no major difference in the behaviors of heat
capacity and compressibility for the systems containing the
protein-like chain.

\subsection{Observed structures and free energy landscape}

\label{sec:subsecenergyland}

Simulations using PT and DMD were performed for three different chain
lengths: $\ell=15,$ 20 and 25, all in a liquid at density
$\rho^*=0.5$. The ranges of temperatures and numbers of replicas
differed in all these cases. The quantities of interest were the
frequencies of occurrence ($f_\mathrm{obs}$) of each
configuration. The most frequently occurring configuration at any
given temperature will be called the dominant configuration if its
population is clearly higher than the second most common structure.

All errors reported below indicate the 95\% confidence intervals,
which is equal to 1.96 times the standard deviation for normally
distributed errors. Below, the configurational potential energy only
refers to the intra-chain bonds energy, and the term \emph{bond}
refers to a hydrogen bond and not the repulsive interactions, unless
otherwise specified.

\subsubsection{15-bead chain}

For $\ell=15$, $95$ temperature values were selected for the replicas,
such that $\Delta\beta_l^*=0.012$ for the highest 10 temperatures and
then for the next 60 temperatures, the $\Delta\beta_l^*$ linearly
decreases to become $0.008$ at the 70th highest temperature and then
$\Delta\beta$ value remains constant for the the rest of temperatures,
while the range of studied temperatures is $T^*_l=[0.76~,~2.5]$
($\beta^*=[1.7~,~5.64]$). The most efficient PT update period was
found in the range of 0.8 to 1.2 picoseconds. Below, the value of 0.8
ps was used. These values are smaller than the PT update period used
in the solvent-less case of~\textbf{I}, which was two picoseconds. As
mentioned above, the number of solvent particles appropriate for this
case to generate the density of 0.5 is $N=1066$, while the length of
the sides of the periodic box is $L=54.4$ \AA.

In Table~\ref{tab:15beads-dominant}, the results for the dominant
configuration at different temperatures are presented for the system
in the presence and in the absence of any solvent.  One sees that at
low enough temperature one structure (BF FJ JN) becomes clearly
dominant as its probability exceeds 60\%.

In Table~\ref{tab:15beads-dominant5.25}, the population and the
average system energies of the most populated structures are provided
for $T_l^*=0.816$ ($\beta^*=5.25$), which is a relatively low
temperature. The configuration with the lowest configurational
potential energy of the system is observed to be the one with the
lowest total potential energy. One also sees that the next three
populated structures in Table~\ref{tab:15beads-dominant5.25} have very
close values for $f_\mathrm{obs}$ and for the average total potential
energy of the system.

\begin{figure}
  \centering
  \includegraphics[width=0.65\columnwidth]{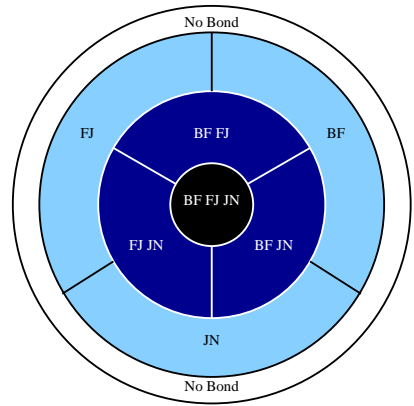}
  \caption{A qualitative picture of the 15-bead chain landscape.}
  \label{Fig:s15-landscape}
\end{figure}

In terms of the free energy landscape, one can interpret these results
as follows. The landscape consists of a relatively deep global minimum
at BF FJ JN, which has three local free energy minima (BF FJ, BF JN,
and FJ JN) close to it, since these configurations differ by only one
bond from the first configuration. Since the last three configurations
in Table~\ref{tab:15beads-dominant5.25} also differ by only two bonds
from the first configuration, their locations in the landscape should
be further from the deepest point such that the configurations 2,3 and
4 should be located between the deepest point and these
configurations. A rough picture of this landscape is presented in
Fig.~\ref{Fig:s15-landscape} in which the distances between the
structures are based on their similarities and the area differences
are related to the differences in their computed entropy in the
absence of the solvent (see I). The lowest free energy structure at
low temperatures, BF FJ JN, has been located in the middle, and the
other structures are positioned based on their similarities to the BF
FJ JN configuration. For example, BF FJ is located between the deepest
point (BF FJ JN) and BF and FJ. This diagram gives some idea about the
folding pathways. For example, to reach the lowest energy structure
with three bonds from the structure with no bonds, initially, the
first bond and then the second bond should be made, which could occur
along six different pathways.

According to Table~\ref{tab:15beads-dominant5.25}, the lowest
potential energy configuration, BF FJ JN, is associated with the
lowest total potential energy of the system as well. However, since
the uncertainty in the computed energies of the 7th structure in
Table~\ref{tab:15beads-dominant5.25} is relatively large, the data in
the table is not sufficient to conclude that the configuration with
the lowest total potential energy is BF FJ JN. However, there are good
arguments for why the first configuration should have the lowest total
potential energy. The first configuration is the most populated one
for all the 66 temperatures that lie in $T^*_l=[0.76~,~2.0]$, so it is
the lowest free energy system at these temperatures. By adding more
bonds and consequently adding more geometrical restrictions, the
configurational entropy decreases, and therefore BF FJ JN has the
lowest configurational entropy among 15-bead configurations. Under the
assumption that the average entropy contributions from the solvent
particles for different configurations are very similar, one can
conclude that the BF FJ JN configuration should have the lowest total
potential energy for $T^*_l=[0.76~,~2.0]$. It is expected that for the
short chains, where the chains do not collapse, the potential energy
difference between two systems mainly depends on their configurational
potential energy difference, while the average potential energy
contribution from the solvent particles will be roughly the same for
different systems energies. To illustrate this, for example, at
$\beta^*=5.25$ ($T_l^*=0.816$) BF FJ JN and BF JN, the two most
populated configurations of Table~\ref{tab:15beads-dominant5.25}, have
on average $0.1 \pm 0.02$ and $0.19 \pm 0.3$ bonds with solvent
particles, respectively. Hence, the contribution to the total
potential energy difference from the bonds between solvent particles
and the chain beads is around 0.02 $\epsilon$, while their
configurational potential energy difference is $1 \epsilon$. The
average number of bonds that each solvent particle makes with other
solvent particles at $\beta^*=5.25$ ($T_l^*=0.816$) is around 5.1.

\begin{table}
  \begin{tabular}{r|l|c|l|c}
    ${\beta^*}$ & in solvent &$f_{\rm obs}$(\%)& solventless& $f_{\rm obs}$(\%)  \\\hline
    1.8 & No bond & 18.2 $\pm$ 0.8 & No bond & 41.3 $\pm$ 1.5 \\
    2.4 & No bond & 18.9 $\pm$0.6 & No bond & 30.15 $\pm$ 1.6  \\
    3.0 & No bond & 11.3 $\pm$ 0.8 & No bond & 19.7 $\pm$ 1.3\\
    3.6 & BF FJ JN & 24.0 $\pm$ 0.9 &BF JN & 17.1 $\pm$ 1.2\\
    4.2 & BF FJ JN & 37.7 $\pm$ 0.8 & BF FJ JN & 26.7 $\pm$ 1.5\\
    4.5 & BF FJ JN & 43.2 $\pm$ 0.9 & BF FJ JN & 35.0 $\pm$ 1.5\\
    4.8 & BF FJ JN & 53.3 $\pm$ 0.8 & BF FJ JN & 44.1 $\pm$ 1.6\\
    5.1 & BF FJ JN & 60.5 $\pm$ 1.2 & BF FJ JN & 55.1$\pm$ 1.8\\
    5.4 & BF FJ JN & 67.3 $\pm$ 0.9 & BF FJ JN & 61.5 $\pm$ 1.6\\
    9 & N/A & N/A & BF FJ JN & 98.6 $\pm$ 0.4 \\
    \hline
  \end{tabular}
  \caption{Most common configurations of the 15-bead chain for
    different temperatures, with and without the solvent.}
  \label{tab:15beads-dominant}
\end{table}

\begin{table}[b]
  \begin{tabular}{r|p{2.0cm}|c|c|c}
    rank  &configuration &$f_{\rm obs}$(\%)&  \multicolumn{2}{c}{potential energy}\\\cline{4-5}
          &                           &      &  total/$\epsilon$&  chain/$\epsilon$ \\\hline
    1 & BF FJ JN & 64.9 $\pm$ 0.9 & -1273.3 $\pm$ 0.3 &-3\\
    2 & BF JN & 10.9 $\pm$ 0.5 & -1271.7 $\pm$ 0.7 & -2\\
    3 & FJ JN & 9.6 $\pm$ 0.5 & -1272.0  $\pm$ 0.7 & -2\\
    4 & BF FJ & 8.9 $\pm$ 0.5 & -1271.8 $\pm$ 0.7 & -2\\
    5 & JN & 1.5 $\pm$ 0.1 & -1271.9$\pm$ 1.8 & -1 \\
    6 & FJ & 1.5 $\pm$ 0.1 & -1272.1 $\pm$ 1.9 & -1 \\
    7 & BF & 1.3 $\pm$ 0.1 & -1272.6 $\pm$ 1.8 & -1\\
    \hline
  \end{tabular}
  \caption{Most populated configurations and their energies of the
    15-bead chain with the solvent environment at $T_l^*=0.816
    (\beta^*=5.25$).}
  \label{tab:15beads-dominant5.25}
\end{table}

\begin{figure}[t]
  \centering
  \includegraphics[angle=-90, width=\columnwidth]{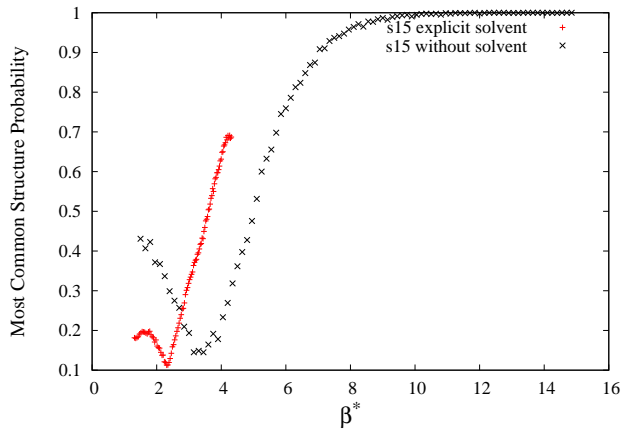}
  \caption{The probability of the most common structure versus inverse
    temperature $\beta^*$ for the 15-bead chain with and without the solvent.}
  \label{fig:s15compared}
\end{figure}

If BF FJ JN has the lowest potential energy of the system, one expects
that the population of this configuration will approach 100\% at lower
temperatures where the free energy mainly depends on the total
potential energy of the system and little on the system entropy. This
trend of the population was indeed seen in the solventless case in
\textbf{I} for chains smaller than 30 beads. The population of the
15-bead dominant structure both with and without the solvent, are
compared in Fig.~\ref{fig:s15compared}, where in both cases (with and
without a solvent) the probability of the dominant structure reaches a
high value. This behavior happens at higher temperatures inside the
solvent, which suggests that the hydrophobic effects of 75\% of the
chain beads assist the folding process and make the helical structures
more favorable. Another consequence of the hydrophobicity is that at
all temperatures, the average radius of gyration for the 15-bead chain
inside the solvent was found to be smaller than in the absence of the
solvent, which shows that the fluid is a poor solvent.

\subsubsection{20-bead chain}

Generally, it is harder to study a wider range of temperatures for
large systems due to the need to use smaller $\Delta \beta$ and
consequently, a higher number of replicas (while each system also
contains more particles). However, as was discussed in
section~\ref{sec:healthyPT}, for the case of a chain in a square-well
solvent, simulating a wide range of temperatures becomes extremely
hard due to the effect of the phase transition on the PT
dynamics. Consequently, for $\ell=20$, the range of studied
temperatures, from $T_l^*=1.05$ to $2.5$ ($\beta^*=[1.7~,~4.1]$), is
smaller than it was for $\ell=15$. To study the energy landscape, 79
temperature values were selected for the replicas such that
$\Delta\beta_l^*=0.008$ for the 40 highest temperatures and then
$\Delta\beta_l^*$ decreases to 0.006 and remains constant for the the
rest of temperatures. For $\ell=20$, the number of appropriate solvent
particles to generate a density of 0.5 for the system with the
periodic box of $L=72$ \AA\ is $N=2522$. The most efficient PT update
period was found to be 320 femtoseconds, which is smaller than the
value of 15-bead case. While it was found that 320 femtoseconds is the
most efficient value for the PT update period of the 20-bead chain
system, there is a range, 260--320 femtoseconds, that yields similar
efficiencies. As mentioned above, around $T_l^*=1.1$, the heat
capacity and compressibility of the fluid show signs of a phase
transition, in line with other studies that predict the location of
the vapor-liquid coexistence line. Because of the phase transition,
trying to reach lower temperatures resulted in poor dynamics even when
using small values for $\Delta \beta^*$, so the lowest temperature was
kept at $T^*_l=1.05$.

The dominant configurations at different temperatures in the presence
and in the absence of the solvent are presented in
Table~\ref{tab:20beads-dominant}. The effective lower limit on
reachable temperatures imposed by the phase transition in the solvent,
made it impossible to determine whether the probability of the most
populated structure approaches one at lower
temperatures. Table~\ref{tab:20beads-dominant3.9} shows that at
$\beta^*=3.9$, BF FJ JN NR is the most populated structure while BF BR
FJ JN NR, which has the lowest configurational potential energy, has a
smaller population. BF FJ JN NR, is a complete helical structure
because it has all the necessary helical bonds between every two
consecutive turns of the protein-like chain; while BF BR FJ JN NR, the
lowest potential energy configuration, is a collapsed helical
structure with an additional bond BR that connects the two ends of the
chain. Hence, BF FJ JN NR, an unfolded but completely helical
structure, has a much higher entropy than the lowest potential energy
configuration, while their energies are close since they only differ
by one internal bond. Furthermore, the complete helical structure can
make more bonds with the solvent particles because of its
non-collapsed shape, as evidenced by the fact that 17\% of the
population of the complete helical structure make bonds with the
solvent particles at $\beta^*=3.9$ ($T_l^*=1.1$), while only 4\% of
the lowest potential energy configuration population make such
bonds. The number of bonds with the solvent particles also shows that
in this model the structures are not soluble, and the potential energy
contribution from bonds between the chain beads and the solvent
particles to the potential energy of the system is relatively small.

\begin{table} 
  \begin{tabular}{r|l|c|l|c}
    ${\beta^*}$ & in solvent &$f_{\rm obs}$(\%)& solventless& $f_{\rm obs}$(\%)  \\\hline
    1.8 & No bond & 7.2 $\pm$ 0.5 & No bond & 28.3 $\pm$ 1.6 \\
    2.4 & No bond & 7.8 $\pm$ 0.5 & No bond & 21.5  $\pm$ 1.3 \\
    3.0 & No bond & 4.6 $\pm$ 0.4 & No bond & 11.6  $\pm$ 1.1\\
    3.3 & BF FJ JN NR & 6.5 $\pm$ 0.5 & BF  & 7.3  $\pm$ 0.9 \\
    3.6 & BF FJ JN NR & 10.3 $\pm$ 0.6 & BF NR & 7.6  $\pm$ 0.8\\
    3.9 & BF FJ JN NR & 12.3 $\pm$ 0.6 & BF JN NR & 9.8 $\pm$ 0.8\\
    4.5 & N/A & N/A &  BF FJ JN NR & 16.3 $\pm$ 1.3 \\
    6.0 & N/A & N/A & BF BR FJ JN NR & 47.7 $\pm$ 1.6 \\
    10.5 & N/A & N/A & BF BR FJ JN NR & 99.1 $\pm$ 0.3 \\
    \hline
  \end{tabular}
  \caption{Most populated configurations of the 20-bead chain inside
    and in the absence of the solvent.}
  \label{tab:20beads-dominant}
\end{table}

According to Table~\ref{tab:20beads-dominant3.9}, the total potential
energies of the most populated structures are very close to each
other. It is therefore hard to predict whether there is a dominant
structure at lower temperatures, as seen in the solvent-less 20-bead
chain.\cite{hanif:34} While the non-collapsed helical structure can
make more bonds with solvent particles in comparison with the lowest
potential energy configuration, the average number of bonds with the
solvent particles is still less than one. It is expected that the
average potential energy contribution from the bonds between solvent
particles is very similar for different configurations. However, it
would require much better sampling statistics than what was obtained
to check this prediction. Since the potential energy of each
intra-chain bond is equivalent to 4.29 bead-solvent bonds, it is
expected that the configuration with the lowest configurational
potential should have the lowest total potential energy as well, and
should, therefore, become the most common structure at lower
temperatures. The reason that the lowest potential energy
configuration does not become dominant at the studied temperatures is
that the BR bond greatly restricts the configurational freedom and
therefore, the non-collapsed helical structure, having one less bond
but with much larger entropy, becomes the most common structure.

Apart from the potential energies and populations of the different
configurations, one can also get an idea of their relative
configurational entropies, although not as precise as in \textbf{I}.
The population of the the lowest potential energy configuration with
the largest number of bonds becomes equal to that of the structure
with no potential energy (no bond) at $\beta^*\approx3.27$, while this
happens at lower temperature, $\beta^*\approx4.05$, in the absence of
a solvent. At $\beta^*=3.27$ in the solvent environment, only 15\% of
the lowest potential energy configurations make bonds with the solvent
particles, while 89\% structures without internal bonds, make bonds
with the solvent particles. By assuming that the potential energy
contribution from bonds between solvent particles is almost the same
for these two configurations, and since they differ by 5 bead-bead
bonds, it can be concluded that the difference in the average
potential energy of the system in this case is likely about
$5\epsilon$ (perhaps somewhat less). The populations of two structures
become equal when their free energy difference is around
zero. Therefore, the entropy difference of the no bond structure and
the lowest potential energy configuration, $\Delta S$, can be
calculated as $\Delta S=5\varepsilon/T=5k_B\beta^*$.  According to
this calculation, in the absence of the solvent $\Delta S \simeq
20.25k_b$ and in the solvent environment $\Delta S \le 16.35 k_b$.
Since these two configurations are the least and the most restricted
ones, respectively, $\Delta S$ represents the maximum configurational
entropy difference. Hence, having a hydrophobic chains in this model
results in a smaller entropy range, which indicates that in comparison
with the previous work in the absence of a solvent, the probability of
the dominant configuration would approach one at higher temperatures,
just as we saw in the previous case of the 15-bead chain,
cf.~Fig.~\ref{fig:s15compared}.

\begin{table} 
  \begin{tabular}{c|p{2.7cm}|c|c|c}
    rank  &configuration &$f_{\rm obs}$(\%)&  \multicolumn{2}{c}{potential energy}\\\cline{4-5}
          &                           &  & total/$\epsilon$&  chain/$\epsilon$  \\\hline
    1 & BF FJ JN NR &12.3 $\pm$ 0.6 & -2482.6 $\pm$ 1.4 & -4\\
    2 & BF FJ NR  & 8.5 $\pm$ 0.4 & -2484.4 $\pm$ 1.6 & -3\\
    3 & BF JN NR & 8.2 $\pm$ 0.4 & -2483.1 $\pm$ 1.6 & -3\\
    4 & BF FJ JN & 7.3 $\pm$ 0.5 & -2482.6 $\pm$ 1.8 & -3\\
    5 & FJ JN NR & 7.3  $\pm$ 0.5 & -2482.7 $\pm$ 1.8 & -3\\
    6 & BF BR FJ JN NR & 5.3 $\pm$ 0.4 & -2483.6 $\pm$ 2.2 & -5\\
    7 & BF JN & 4.6 $\pm$ 0.4 & -2484.1 $\pm$ 2.2 & -2\\
    \hline
  \end{tabular}
  \caption{Most populated configurations of the 20-bead chain inside
    the solvent at $\beta^*=3.9$ ($T_l^*=1.1$).}
  \label{tab:20beads-dominant3.9}
\end{table}

\subsubsection{25-bead chain}

For the 25-bead chain, the system needs to include 4644 solvent
particles, which is nearly twice the number of solvent particles as in
the 20-bead system, in a box of $L=88$ \AA. According to
Fig.~\ref{fig:HCaverage}, the temperature at which the phase
transition behavior is observed increases slightly with increasing
$N$. Therefore, the range of temperatures that could be investigated
for the 25-bead chain system is even smaller than 20-bead case. A set
of temperatures with 95 replicas was chosen, such that for the 20
highest temperatures $\Delta\beta=0.006$, and for the rest of
temperatures $\Delta\beta=0.004$, while the range of studied
temperatures is $T^*_l=[1.22~,~2.5]$ ($\beta^*=[1.7~,~3.5]$). The most
efficient PT update period is 120 femtoseconds, which is even smaller
than in the 20-bead case.

According to Table~\ref{tab:25beads-dominant}, the structure with the
lowest configurational potential energy becomes dominant at higher
temperatures in comparison with the solvent-less case of
\textbf{I}. However, the range of studied temperature is not
sufficient to observe the kind of deep funnel in the free energy
landscape at low temperatures that was observed in the absence of a
solvent. The most common structures of the 25-bead chain inside a
solvent at $\beta^*=3.3$ are presented in
Table~\ref{tab:25beads-beta3.3}. While the populations of
configurations 3-10 are equal within the statistical error, the
population of the first configuration (with 8 bonds) is clearly higher
than that of the other configurations. Our study reveals that this
configuration is clearly the most populated one for $T_l^* \le 1.32$
($\beta^* \ge 3.24$). This means that for all the temperatures in the
range $1.22 \le T_l^* \le 1.32$ ($3.24\ge\beta^*\ge 3.0$), the
structure with the lowest configurational potential energy is the most
common structure.  Since the configurational entropy decreases by
increasing the number of bonds (because of adding more restrictions),
the first configuration should have the lowest configurational
entropy.  Since the first configuration has been the most common
structure at the lowest studied temperature, the system containing the
first configuration should be the lowest total potential energy at
these temperatures. It is expected that by decreasing the temperature,
the order of the system energies does not change dramatically and
therefore, when decreasing the temperature, the first configuration
likely remains the one with the lowest total potential energy and
therefore, the population of this structure should approach one at low
temperatures, similar to the results of \textbf{I}.

For the 25-bead case, a similar reasoning as for the 20-bead chain
leads to the prediction of a large configurational entropy difference
between the lowest potential energy configuration with a completely
collapsed shape (first configuration of
Table~\ref{tab:25beads-beta3.3}) and the non-collapsed helical
structure (second configuration of
Table~\ref{tab:25beads-beta3.3}). Therefore, one anticipates that the
non-collapsed helical structure has the highest occupancy for the
limited range of temperatures that was studied in the
simulations. However, because of the potential energy difference of
3$\epsilon$, the first configuration becomes dominant, even at not
very low temperatures. This is unlike the 20-bead chain, for which the
non-collapsed helical structure (1st configuration of
Table~\ref{tab:20beads-dominant3.9}) is dominant at similar
temperatures. Consequently, the probability of the collapsed helical
structure of the 25-bead chain approaches one at lower temperatures as
it does in the absence of the solvent.

\begin{table}
  \begin{tabular}{r|p{2.4cm}|c|p{2.4cm}|c}
    ${\beta^*}$ & in solvent &$f_{\rm obs}$(\%)& solventless& $f_{\rm obs}$(\%)  \\\hline
    1.8 & No bond & 2.7 $\pm$ 0.3 & No bond & 21.7 $\pm$ 1.3 \\
    2.4 & No bond & 3.3 $\pm$ 0.3 & No bond &  15.5$ \pm$ 1.0  \\
    3.0 & NR & 1.3 $\pm$ 0.2 & No bond &  6.7 $\pm$ 0.9 \\
    3.3 & BF BR BV FJ FV JN NR RV & 2.6 $\pm$ 0.2 & No bond &4.3 $\pm$ 0.6\\
    4.5 & N/A & N/A & BF BR BV FJ FV JN NR RV & 7.5 $\pm$ 1.0\\
    9.0 & N/A & N/A& BF BR BV FJ FV JN NR RV & 98.0 $\pm$ 0.4\\
    \hline
  \end{tabular}
  \caption{Most populated configurations of the 25-bead chain inside
    and in the absence of the solvent.}
  \label{tab:25beads-dominant}
\end{table}

\begin{table}
  \begin{tabular}{r|p{3.1cm}|c|c|c}
    rank  &configuration &$f_{\rm obs}$(\%)&  \multicolumn{2}{c}{potential energy}\\\cline{4-5}
          &                           &      &  total&  chain  \\\hline
    1 & BF BR BV FJ FV JN NR RV & 3.4 $\pm$ 0.3 & -4219.9 $\pm 1.9$ & -8 \\
    2 & BF FJ JN NR RV &  2.0 $\pm$ 0.3 & -4217.7 $\pm$ 2.5 & -5 \\
    3 & FJ JN NR RV & 1.5 $\pm$ 0.2 & -4216.0 $\pm$ 2.7 & -4 \\
    4 & BF FJ JN NR & 1.5 $\pm$ 0.3 &-4219.4 $\pm$ 2.9 & -4 \\
    5 & BF FJ JN RV & 1.5 $\pm$ 0.3  & -4215.8 $\pm$ 3.0 & -4 \\
    6 & BF BR BV FJ JN NR RV & 1.3 $\pm$ 0.2 & -4218.4 $\pm$ 2.9 & -7 \\
    7 & BF FJ NR RV & 1.3 $\pm$ 0.2  &-4215.2 $\pm$ 3.0 & -4 \\
    8 & BF JN NR & 1.3 $\pm$ 0.1 & -4213.0 $\pm$ 3.7 & -3 \\
    9 & JN NR RV & 1.2 $\pm$ 0.1 & -4216.3 $\pm$ 2.9 & -3 \\
    10 & FJ JN RV & 1.2 $\pm$ 0.1  & -4213.1 $\pm$ 3.1 & -3 \\
    \hline
  \end{tabular}
  \caption{Most populated configurations of the 25-bead chain inside
    the solvent at $\beta^*=3.3$ ($T_l^*=1.30$).}
  \label{tab:25beads-beta3.3}
\end{table}

\section{Conclusions}
\label{sec:concl}

The free energies of different configurations (i.e., the free energy
landscape) of a protein-like chain in a solvent at different
temperatures were investigated. Qualitatively, the behavior of a
protein-like chain inside a square-well solvent is similar to the
behavior in the absence of a solvent, studied in \textbf{I}. For the
15-bead chain, the lowest free energy configuration was found to be an
alpha helix that becomes dominant at low temperatures, just as it did
in the absence of any solvent. The free energy landscape of the
15-bead chain at low temperatures consists of a funnel with a very
deep global minimum and a few local minima around it. By lowering the
temperature, the global minimum becomes deeper while the others become
shallower and consequently, the funnel becomes steeper.

For larger chain lengths, in particular, for $\ell=20$ and $\ell=25$,
a phase transition of the square-well solvent effectively puts a lower
bound on the temperature range accessible in the simulations. The
observed phase transition temperature coincides roughly with the
temperature at which previous studies observed a liquid-vapor
coexistence line. Investigating the free energy landscape of a
solvated system over a phase transition point of the solvent can be
very challenging using the PT method, especially for larger
systems. For the 20-bead and 25-bead chains the effects of the phase
transition become more apparent because of the larger number of
particles in comparison with the 15-bead chain. Consequently, the
temperature range studied here could not be extended below the
(effective) phase transition temperature for the 20-bead and 25-bead
chains. This difficulty is not easy to overcome, since it is related
to the efficiency of the PT algorithm itself near the phase transition
point. Substantial computational resources, over a million cpu hours,
were used to obtain the results presented here, which were mainly
utilized to obtain the best set of parameters for the PT runs. As a
result of the considerable computational demand of computing the free
energy of the solvated system below the phase transition point, a
direct comparison for the 20-bead and 25-bead chain systems with the
previous study could not be done for the whole range of temperatures
that were investigated in \textbf{I}.  However, it is expected that
for both 20-bead and 25-bead chain systems the configuration with the
lowest configurational energy becomes dominant at lower temperatures,
since their systems energy seem to be the lowest ones at very low
temperatures, which is mainly due to their low configurational energy
and the hydrophobicity of 75\% of protein-like chain beads (having
only hard-core repulsive interactions with solvent particles).

While for the 15-bead chain the lowest potential energy configuration
is an unfolded alpha-helix without any specific tertiary structure,
for longer chains, the bonds between different layers of the helix,
such as the bond between two ends of the chain, cause the lowest
potential energy structure to be a folded structure. Our study showed
that the entropic barrier for making bonds between the two ends of the
chain for longer chains is larger than the change in entropy
associated with forming a helix structure. As a consequence, the
unfolded helical structure is dominant for a relatively large range of
temperature until the low-temperature regime where the folded helix is
favored. Therefore, similar to the absence of a solvent, the effect of
temperature on the morphology of the landscape is more apparent for
the longer chains.

One of the major differences between a protein model in a solvent
(studied here) and without a solvent (studied in \textbf{I}) is the
effect of mainly repulsive interactions of the beads with the solvent
particles in the folding process. Only 25\% of the beads can make
attractive bonds with the solvent, while the rest of the beads only
have repulsive interactions with the solvent (i.e., they are
hydrophobic). Because of the restriction effects of the repulsive
interactions, the entropy range (i.e., the entropy difference of the
minimum and maximum number of bonds configurations) is smaller in
comparison with the absence of a solvent. Because of the smaller
entropy range, the landscape shows funnel behavior at higher
temperatures in comparison with the absence of a solvent.

The main problem in studying the protein-like chain inside the solvent
is the slow convergence of estimates of the free energy of
configurations using the PT method. For example, the presence of a
phase transition in the square-well fluids leads to large statistical
errors in the PT method. To overcome the effects of the phase
transition on sampling, the PT method should be enhanced by
incorporating other techniques, such as the umbrella
sampling.\cite{Valleau:48} Another solution for this problem is to use
different parameters for the square-well liquid, such that the phase
transition temperature lies outside the temperature range of
interest. According to Ref.~\onlinecite{OreaRf4:36}, by increasing the
ratio $\lambda=\sigma'/\sigma$, the liquid-vapor coexistence line
shifts to higher temperatures for the density of $\rho^*=0.5$. For
example, for $\lambda=2.0$ the liquid-vapor coexistence line is
crossed at a temperature around $T_l^*=2.4$ for
$\rho^*=0.5$,\cite{OreaRf4:36, ErringtonRf3:35, Miguel:49} which is
very close to the highest studied temperature ($T_l^*=2.5$). While
using $\lambda \ge 2$ can be helpful for avoiding the phase
transition, it would allow for an unphysically large range of bond
vibrations in comparison with real proteins.

One of the possible avenues for future research is to investigate the
dynamics of the folding transition, instead of only the resulting free
energies.  While studying the folding pathways can be computationally
very demanding, it provides more information about the nature
of~folding.  Some earlier studies have provided some simple
connections between energy landscapes and protein folding kinetics,
which can be applied (under suitable assumptions) to this
study.\cite{Szabo:77, Hamm:78} For example, by considering the
distance between any two beads that can make a bond as a reaction
coordinate, it is possible to observe the potential of mean force
(Helmholtz free energy) versus the reaction coordinate (distance of
the two beads that can make a bond). From this potential of mean
force, one can estimate first passage times (Kramers' problem) to
extract rates of forming and breaking a bond. These rate enter the
relaxation matrix in a rate equation approach, which could then be
used to analyze to the dynamics of protein folding.

\begin{acknowledgments}
  Computations were performed on the GPC supercomputer at the SciNet
  HPC Consortium, which is funded by the Canada Foundation for
  Innovation under the auspices of Compute Canada, the Government of
  Ontario, the Ontario Research Fund Research Excellence and the
  University of Toronto. This work was supported by a grant from the
  Natural Sciences and Engineering Research Council of Canada.
\end{acknowledgments}

\end{document}